\documentclass[useAMS,usenatbib]{mn2e}
\usepackage[letterpaper,margin=0.65in]{geometry}
\usepackage[ansinew]{inputenc}
\usepackage{amsfonts}
\usepackage{amsmath}
\usepackage{graphicx}
\usepackage{epsfig}
\usepackage{color}
\usepackage{amssymb}
\usepackage[normalem]{ulem}

\newcommand{\eg}{{e.g., }}
\newcommand{\ie}{{i.e., }}
\newcommand{\pz}{photo-$z$\ }
\newcommand{\pzs}{photo-$z$s }
\newcommand{\tpz}{\texttt{TPZ} }
\newcommand{\tpzns}{\texttt{TPZ}}

\newcommand{\somzns}{\texttt{SOM$z$}}

\newcommand{\pzns}{photo-$z$} 
\newcommand{\pzsns}{photo-$z$s} 
\newcommand{\PZ}{Photo-$z$\ }

\renewcommand{\vec}[1]{\mathbf{#1}}
\DeclareMathOperator\erf{erf}

\title[Sparse representation of \pz PDFs]
{Sparse Representation of Photometric Redshift PDFs: Preparing for Petascale Astronomy}
\author[M. Carrasco Kind and R. J. Brunner] 
{Matias Carrasco Kind\thanks{E-mail: mcarras2@illinois.edu} and Robert J. Brunner\\
Department of  Astronomy, University of Illinois, Urbana, IL 61820 USA}

\begin{document}
\date{\today}

\pagerange{\pageref{firstpage}--\pageref{lastpage}} \pubyear{2014}

\maketitle

\label{firstpage}
\begin{abstract}
One of the consequences of entering the era of precision cosmology is the widespread adoption of photometric redshift probability density functions (PDFs). 
Both current and future photometric surveys are expected to obtain images of billions of distinct galaxies. As a result, storing and analyzing all of these PDFs will be non-trivial and even more severe if a survey plans to compute and store multiple different PDFs. In this paper we propose the use of a sparse basis representation to fully represent individual photo-$z$ PDFs. By using an Orthogonal Matching Pursuit algorithm and a combination of Gaussian and Voigt basis functions, we demonstrate how our approach is superior to a multi-Gaussian fitting, as we require approximately half of the parameters for the same fitting accuracy with the additional advantage that an entire PDF can be stored by using a 4-byte integer per basis function, and we can achieve better accuracy by increasing the number of bases. By using data from the CFHTLenS, we demonstrate that only ten to twenty points per galaxy are sufficient to reconstruct both the individual PDFs and the ensemble redshift distribution, $N(z)$, to an 
accuracy of 99.9\% when compared to the one built using the original  PDFs computed with a resolution of $\delta z = 0.01$, reducing the required storage of two hundred original values by a factor of ten to twenty. Finally, we demonstrate how this basis representation can be directly extended to a cosmological analysis, thereby increasing computational performance without losing resolution nor accuracy.
\end{abstract}


\section{Introduction} 
Over the last two decades, photometric redshift (hereafter \pzns) estimation has become a crucial requirement for modern, multi-band photometric galaxy surveys. The need for accurate and robust \pzsns, and in particular for their probability density functions (\pz PDFs), is increasing at an even faster rate as we enter the era of precision cosmology.  Large photometric surveys like the Dark Energy Survey (DES\footnote{http://www.darkenergysurvey.org/})  and the Large Synoptic Survey Telescope (LSST\footnote{http://www.lsst.org/lsst/}) will probe hundreds of millions to billions of galaxies, which are often too faint to be spectroscopically observed. Even today, the Sloan Digital Sky Survey (SDSS;~\citealt{York2000}) has obtained hundreds of millions of photometric observations of extragalactic sources covering approximately one-third of the sky. This same survey has also, by using a considerably larger amount of time with the same telescope, obtained a galaxy spectroscopic sample that is about one hundred 
times smaller. While the spectroscopic sample is a higher precision sample~\citep{Ahn2013}, spectroscopy is both considerably more difficult and more time consuming than photometry.  For a number of cosmological analyses this higher precision is not strictly required, thus photometric surveys have become a cost-effective method for enabling new cosmological constraints.

As these photometric surveys have grown in popularity, the estimation of \pzs that use different techniques has also grown significantly. These techniques can be broadly separated into machine learning algorithms~\citep[\eg][]{Collister2004, CarrascoKind2013}  or Spectral Energy Distribution (SED) fitting methods~\citep[see \eg][]{Benitez2000,Ilbert2006}. These two classes of techniques have been shown to have similar performance characteristics when high quality training data are available~\citep[see, \eg][for a review on some current \pz techniques]{Hildebrandt2010,Abdalla2011,Sanchez2014}. More recently, particular attention has been focused on techniques that compute a full \pz PDF for each galaxy. This is because a \pz PDF contains more information than a single \pz estimate, and the use of \pz PDFs has been shown to improve the accuracy of cosmological measurements~\citep[\eg][]{Mandelbaum2008,Myers2009,Sheldon2012,Carnero2012,Jee2013} while not introducing any biases~\citep[\eg][]{Bordoloi2010,
Abrahamse2011}.

One fact that all photometric surveys have in common is the need to efficiently handle an overwhelming quantity of imaging data. The reduction, analysis and storage of this data is a difficult problem; even with the growth of computational resources, efficiently handling these data remains a pressing problem. In particular, \pz PDFs are currently computed on the summary catalogs that are produced by uniformly processing imaging data.  But storing \pz PDFs for billions of sources is a challenge in itself, which is further complicated if multiple, different \pz techniques are desired or if different \pz PDFs are generated by using different galaxy templates. This is a problem both for those managing the data archives and for the general community who desire to apply these \pz PDF estimates to cosmological analyses. Thus the time is ripe to address this issue. In this paper, therefore, we explore different methods that allow us to manipulate and use \pz PDFs in a more efficient manner by representing them as 
compactly as possible.

In this paper we introduce the use of a sparse functional basis to represent a full \pz PDF. This approach minimizes the data required to represent the \pz PDF, while maximizing the accuracy of the PDF. This basis representation not only minimizes the storage requirements, but also allow us to manipulate PDFs in a more computationally efficient manner, thereby increasing the computational efficiency of resulting analyses. With this approach, each galaxy \pz PDF is decomposed into an over determined basis system by minimizing the number of basis functions retained. We analyze how this approach compares with other representation techniques, in particular with a multi-Gaussian approach; and we demonstrate that, by using our proposed functional form, the integration and manipulation of \pz PDFs is both easier and faster than when using either the original PDF or any other comparable technique.

The rest of this paper is organized in the following manner. In Section 2 we first present the data we will use to test our proposed approach, after which we will introduce the \pz methods that we will use to generate the \pz PDFs that will be used in the rest of the paper. Section 3 describes the different algorithms used to represent the \pz PDFs. In Section 4 we present the main results, compare the performance of the different representation methods, and discuss the storage requirements of these \pz PDFs. Next, we apply our basis representation to the computation of $N(z)$ in Section 5, and discuss how this basis framework can reduce computational costs. We conclude in Section 6 with a summary of our main findings and a final discussion of the applications of our proposed approach.

\section{Data and photo-z estimation}\label{data}
In this section, we first present a brief summary of the data  that will be used to test different \pz PDF representation techniques, including our proposed sparse representation basis. Next, we will present a short overview of \tpz, the \pz PDF technique we use to generate the \pz PDFs that are analyzed in this paper.

\begin{figure}
\includegraphics[width=0.45\textwidth]{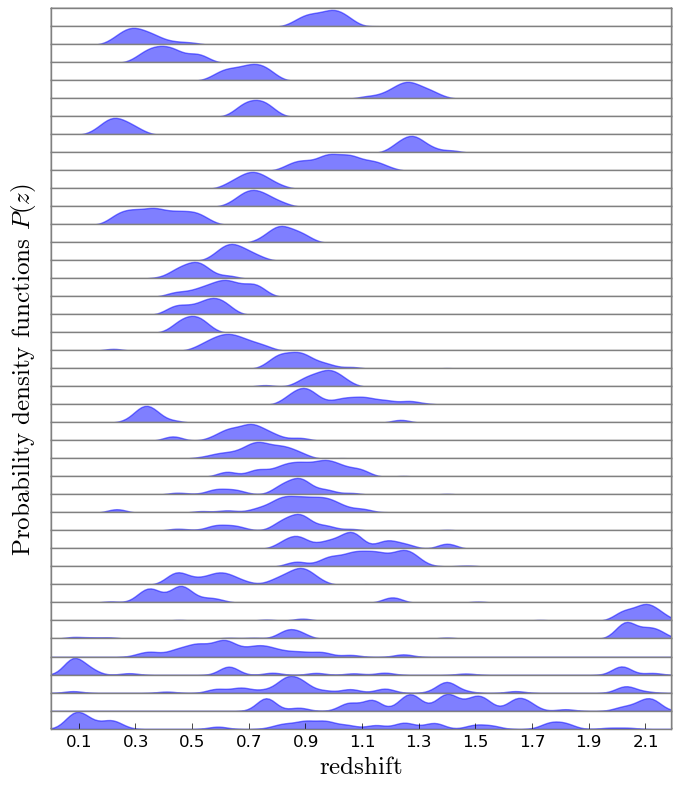}
\caption{Forty representative, randomly selected \pz PDFs computed for the CFHTLenS data by using \tpzns, each normalized to unity. In each subplot, the horizontal axis is redshift and the vertical axis is the probability density. The PDFs are sorted in the vertical axis by the number of peaks.}
\label{fig:36pdf}
\end{figure}

\subsection{CFHTLenS}

We use data from the Canada-France-Hawaii Telescope Lensing Survey~\citep{Heymans2012,Erben2013}, hereafter referred to as CFHTLenS\footnote{http://www.cfhtlens.org/}, with the photometry presented in \cite{Hildebrandt2012}. This galaxy survey includes more than twenty-five million objects observed in five photometric bands:  $u$, $g$, $r$, $i$, and $z$, covering 154 square degrees that includes all five years worth of data from the Wide, Deep and Pre-survey components of the CFHT Legacy Survey~\citep[CFHTLS;][]{Gwyn2012}.  To generate a spectroscopic training sample, we have cross matched the galaxies from the CFHTLenS with spectroscopic surveys whose sky coverage overlaps the four fields of the CFHTLenS survey. 

In particular, we have selected high quality spectroscopic galaxies from the  Deep Extragalactic Evolutionary Probe Phase 2~\citep[DEEP2;][]{Davis2007,Newman2013a}, the VIMOS (VIsible imaging Multi-Object Spectrograph ) VLT (Very Large Telescope) Deep Survey~\citep[VVDS;][]{LeFevre2005,Garilli2008}, the VIMOS 
Public Extragalactic Redshift Survey~\citep[VIPERS;][]{Garilli2014}, and the Sloan Digital Sky Survey Data Release 10~\citep[SDSS-DR10]{Ahn2013}, which includes over two million spectra of galaxies and quasars taken as a part of the the Baryonic Oscillation Spectroscopic Survey (BOSS) program \citep{Dawson2013}. In the end, we have 49,868 high quality spectroscopic galaxies with a mean redshift of 0.6 to train our \tpz algorithm. As the objective of this paper is not to present photometric redshift PDFs for all of the CFHTLenS galaxies, we have randomly selected a subsample of $10^6$ galaxies with no spectroscopic information from the survey that we use for the tests described within this text. 

\subsection{TPZ: Trees for Photo-$Z$}
\tpz \citep{CarrascoKind2013} is a parallel, supervised machine learning algorithm that uses prediction trees and random forest techniques to produce both robust \pz PDFs and ancillary information for a target galaxy sample. A prediction tree is built by \textit{asking} a sequence of questions of the data. This process recursively splits the input data taken from the spectroscopic sample, frequently into two branches, until a terminal leaf is created that meets a stopping criterion (\eg a minimum leaf size or a variance threshold). The dimension in which the data is divided is chosen to be the one with highest information gain  among the random subsample of dimensions obtained at every point. This process produces less correlated trees and allows us to explore several configurations within the data. The small region bounding the data in the terminal leaf node represents a specific subsample of the entire data with similar properties. Within this leaf, a model is applied that provides a fairly comprehensible 
prediction, especially in situations where many variables may exist that interact in a nonlinear manner, as is often the case with \pz estimation. 

\tpz is publicly available through a software package called MLZ\footnote{http://lcdm.astro.illinois.edu/code/mlz.html} (Machine Learning for photo-Z). The full software package includes, among other capabilities,  \tpzns, a supervised learning method, and \somzns, an unsupervised \pz technique \citep{CarrascoKind2014a}. As we have demonstrated in ~\citet{CarrascoKind2014b}, these techniques can be combined to make even more accurate \pz PDF predictions. For simplicity, in this paper we restrict our analysis to \tpz only, since we focus within this paper solely on the accurate reconstruction of galaxy \pz PDFs.

We compute the \pz PDF for all one million galaxies in our test sample, described previously, by using our spectroscopic training sample. We used all colors and magnitudes, which results in a total of nine attributes, and construct 600 trees to make the predictions. \tpz also uses the attribute errors during the prediction process, in part to deal with missing attributes in the catalog~\cite[see][for a description of this approach]{CarrascoKind2013}. We also have computed photometric redshifts for galaxies in the training sample by using a cross validation technique called Out-Of-Bag~\citep{Breiman2001,CarrascoKind2013} in which a \pz PDF is obtained for all galaxies in the training set by using all the trees that do not contain that particular galaxy. This approach, therefore, avoids over-fitting; and we have shown that this method is reliable and also unbiased~\citep{CarrascoKind2014a}. 

\begin{figure}
\includegraphics[width=0.45\textwidth]{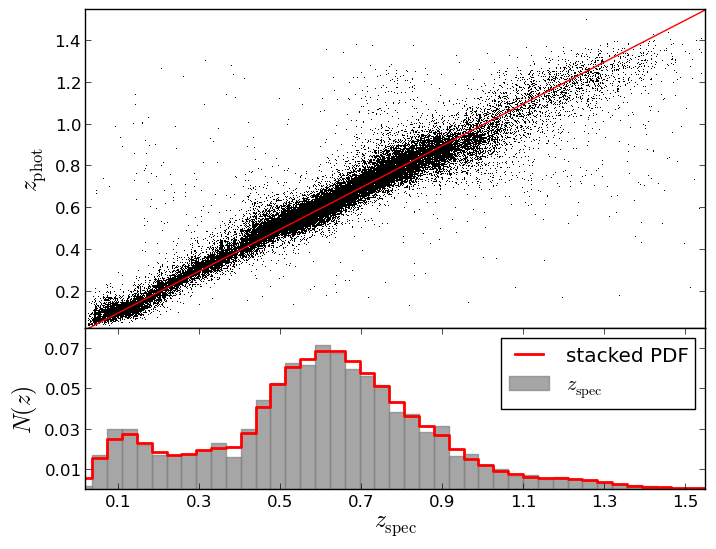}
\caption{\textit{Top:} A comparison of the photometric and spectroscopic redshift for all galaxies in the training sample computed by using the OOB cross-validation technique. The red line shows the one-to-one line for guidance. \textit{Bottom:} The $N(z)$ distribution computed by using the spectroscopic redshifts (gray area) and by stacking the \pz PDFs (red) for the training sample galaxies.}
\label{fig:oob_nz}
\end{figure}

For illustration, we present a sample of forty \pz PDFs randomly selected from the CFHTLenS galaxies in Figure~\ref{fig:36pdf}, presented in increasing order by the computed mean value of their \pz PDF. The redshift range for the galaxies are the same and the PDFs have all been normalized to unity. From this figure, it is clear that these \pz PDFs are not simple functions, often having multiple peaks; and they are, therefore, poorly represented by a single Gaussian, which has often been used for simplicity in the past. In Figure \ref{fig:oob_nz}, we present a summary of the results on the training data determined by using the OOB cross-validation technique. The top panel compares $z_{\rm phot}$, computed by using the mean value of each \pz PDF, with $z_{\rm spec}$ for all  49,868 galaxies in the training sample. This indicates the approximate performance of \tpz on the real dataset within the limits of the training sample. The bottom panel shows the distribution of the galaxies as a function of redshift 
in terms of their spectroscopic values (in gray) compared with the $N(z)$ obtained by stacking the PDFs (red line). We can see a remarkable agreement between these two distributions, as we have shown previously~\citep{CarrascoKind2013}.

\section{PDF Representation}

In this section we present the different methods that we use to represent the full \pz PDF. For the rest of this discussion, we will make the following assumptions. First, we have $N$ total galaxies in our sample with individual \pz PDF estimates. Second, we can represent the \pz PDF, $P_k(z)$, for the $k^{\textrm{th}}$ galaxy in the sample by $\vec{pz}_k$. Finally, we have $n$ sample points in the original galaxy \pz PDF. Thus $P_k(z)$ is sampled at a resolution $\delta z$, given by $\delta z = \Delta z / n$, where $\Delta z$ is the redshift range spanned by the photometric data.   

\subsection{Statistical Representation}\label{singlez}

The simplest representation for a full \pz PDF is to use a summary statistic. We will consider five different, summary statistics, four of which are single values: the mean, the mode, the median of the PDF, and a Monte-Carlo sampling from the original cumulative PDF~\citep{Wittman2009}. This fourth approach involves sampling a random number from the range $0$--$1$, and determining the cumulative probability to this numerical value. Finally, the last representation is a single Gaussian fit to the original \pz PDF, which provides a two value summary: the mean and variance of the Gaussian.  As a result, we require either $N$, for the first four approaches, or $2N$, for the last approach, statistics to represent the full \pz PDF catalog.

\subsection{Multi-Gaussian Fitting}

The second representation for a full \pz PDF we explore is the application of a multi-Gaussian fit to each \pz PDF~\citep[see, \eg][]{Bovy2011,Bovy2012}. In this approach, each Gaussian (when more than one is used) included during the fitting process will require three parameters: the amplitude, the mean, and the variance. In this approach, we first determine the number of peaks, $Np_k$, in the \pz PDF. We increase this value by one and use the result as the number of Gaussians to be used in the fitting process for that specific \pz PDF. The extra Gaussian improves the fit to extended wings in the \pz PDF distribution, which often arise from the residuals of the Gaussian fits to the individual PDF peaks.

To determine the best fit values, we use a Levenberg-Marquardt minimization algorithm. In this case, each $P_k(z)$ can be represented by:
\begin{equation}
 \vec{pz}_k = \sum\limits_{i=1}^{Np_k+1} \alpha_{k,i} e^{-\frac{(\vec{z}-\mu_{k,i})^2}{2\sigma_{k,i}^2}}
\end{equation}
where $\boldsymbol{\alpha}_k$, $\boldsymbol{\mu}_k$ and $\boldsymbol{\sigma}_k^2$ are vectors of dimension $Np_k+1$ containing the amplitude, mean, and variance for each Gaussian included in the fitting process. As a result, the total number of values needed to represent the full \pz PDF catalog is $\sum_k 3(Np_k+1)$.

\subsection{Sparse Basis Representation}

The final technique that we will use to represent a \pz PDF is the sparse basis representation. In this case, we will adopt a set of basis functions to represent a PDF by using the following model:
\begin{equation}
 \vec{pz}_k = \vec{D} \boldsymbol{\delta}_k + \boldsymbol{\epsilon}_k
\end{equation}
where $\vec{D}$  is a dictionary or basis matrix of dimension $n \times m$, where $m > n$. Thus, we have an over-determined problem as the number of basis functions, $m$, is much larger than the dimension, $n$,  of each \pz PDF. In this case, each column, $\vec{d}_j$, of the dictionary matrix, $\vec{D}$, is a basis function that must be $\ell_2$ normalized, \ie $|| \vec{d}_j ||_2 =1$ for $j = 1,2,\dots, m$. 

We want to find, for each galaxy $k$, the optimal vector solution $\boldsymbol{\delta}_k$, which is determined such that its pseudo-norm $||\boldsymbol{\delta}_k||_0$ is minimized. Alternatively, this can be equivalently stated that we want to minimize the number of non-zero entries in the vector, $\boldsymbol{\delta}_k$, given the residual error $\boldsymbol{\epsilon}_k$. In this case, we can either use a predefined number of basis functions or we can define a fixed residual for every galaxy in the sample. Either way, the total number of points required to represent the entire catalog is given by $\sum_k 2(Nb_k)$, where $Nb_k$ is the number of basis functions used for each galaxy. Note that in this case we only need two numbers for each functional basis: the functional coefficient (a floating point number), and the index number of the function within the basis set (a integer number). This already corresponds to a potentially large reduction in the total data volume required to archive \pz PDFs. We will see 
in subsequent sections that we can also represent the basis function coefficients by using integers; and that, in addition, we can combine both terms into a single thirty-two bit integer, thereby reducing the total number of values required to $\sum_k (Nb_k)$.

Finding $\boldsymbol{\delta}_k$ in this over determined problem can be challenging. For this analysis, we have selected to use Orthogonal Matching Pursuit (OMP), an iterative algorithm that finds, at each step, the column, $\vec{d}_j$,  of the dictionary matrix, $\vec{D}$, that best represents the current residuals. This process is repeated until a predefined criteria is reached, either a residual threshold or the total number of basis functions used. Fundamentally, this approach is similar to the well known \texttt{CLEAN} algorithm, which is used to analyze interferometric radio observations~\citep{Hogbom1974}. The advantage of  OMP over the standard Matching Pursuit algorithm~\citep{Mallat1993} is that a specific basis function can only be selected once. Since the residuals are orthogonalized during the selection of the basis functions for the current galaxy, we generate an independent set of basis functions to represent each galaxy's \pz PDF. 

Conceptually, the OMP algorithm that we apply to all galaxies can be enumerated\footnote{Note, the superscript $T$ indicates transposition}:
\begin{enumerate}
\renewcommand{\theenumi}{\arabic{enumi})}

\item Initialize all variables. First, define the residual vector to be the original \pz PDF, $\boldsymbol{\epsilon}_k^{0} = \vec{pz}_k$. Second, create an empty set of cumulative selected basis functions, $\vec{B}_k$. Finally, set $\boldsymbol{\delta}_k = 0$, and define $i=0$ as the number of the current iteration.

\item Compute the current set of basis functions. First, find the column vector, $\vec{d}_b$, from the dictionary matrix, $\vec{D}$, where $b$ is the index position that maximizes the projection of $\boldsymbol{\epsilon}_k^i$:
\begin{equation}
\vec{d}_b^i = \max_{\vec{d}_j \in \vec{D}} | \vec{d}_j^T \cdot \boldsymbol{\epsilon}_k^{i} | 
\end{equation}
Second, add this selected basis function to the set $\vec{B}_k$, \ie $\vec{B}_k = (\vec{B}_k,\vec{d}_b^i)$.

\item Orthogonally project the original \pz PDF onto the linear space spanned by the columns of all previously selected basis functions:
\begin{equation}
\vec{w}_k^i = \vec{B}_k^T \cdot \vec{pz}_k 
\end{equation}
where $\vec{w}_k^i$ is a temporary vector corresponding to the coefficients of the currently used basis functions in $\vec{B}_k$.

\item Complete the projection by updating the residuals by using the temporary vector $\vec{w}_k^i$:
\begin{equation}
\boldsymbol{\epsilon}_k^{i+1} = \vec{pz}_k - \vec{B}_k \cdot \vec{w}_k^i
\end{equation}

\item Check the stopping criteria: $||\boldsymbol{\epsilon}_k^{i+1} ||_2 < \epsilon_{th}$, where $\epsilon_{th}$ is the threshold residual or  $i > i_{lim}$, where $i_{lim}$ is the number of required basis functions. If the pre-selected stopping criteria is met, the calculations are completed: $\boldsymbol{\delta}_k = \vec{w}_k^i$ and $\vec{pz}_k = \vec{D} \cdot \boldsymbol{\delta}_k + \boldsymbol{\epsilon}_k^{i+1}$, where $\boldsymbol{\delta}_k$ is sparse. Finally, the \pz PDF representation is defined:
\begin{equation}
\vec{pz}_k \approx \vec{D} \cdot \boldsymbol{\delta}_k
\end{equation}
On the other hand, if the predefined stopping criteria is not met, the iteration step is increased, $i = i + 1$, and steps $2$--$5$ are repeated by using the current residual vector. This process is repeated over all galaxies $k$, where $k=1,2,\dots,N$.
\end{enumerate}

\begin{figure}
\includegraphics[width=0.48\textwidth]{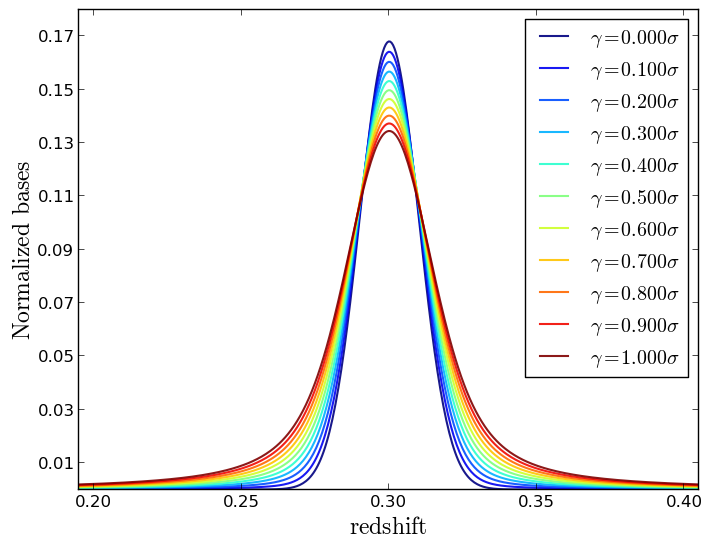}
\caption{Different normalized $||\vec{d}_j||_2=1$ Voigt profile basis functions with the same mean, $\mu = 0.3$, and sigma, $\sigma=0.01$, for different values of $\gamma$, which ranges from $0$ (blue) to $1\sigma$ (red). Note that for $\gamma = 0$, we recover the standard Gaussian distribution. In a full dictionary, we create these profiles over the entire redshift range of the galaxy sample for different values of $\sigma$.}
\label{fig:voigt}
\end{figure}

\subsubsection*{Dictionary Selection}
Given the nature of the shape of the \pz PDFs (see, \eg Figure~\ref{fig:36pdf}), it is natural to select a set of Gaussian-like basis functions that span the redshift range of our photometric galaxy sample. We can use the the original resolution and redshift range spanned by the generated \pz PDF to determine the dictionary to use for the sparse basis representation. One of the primary advantages of this method is that these dictionary entries are composed of analytic functions that can be combined with other functional forms. There are no restrictions, other than computational time, on how large of a dictionary we can use,  as there is no requirement for the dictionary to be permanently stored. Furthermore, a \pz PDF can be restored even without reconstructing the dictionary, as long as the indices and coefficients are efficiently stored. 

We select $N_{\mu}$ Gaussian functions,  whose mean values span the redshift range of our galaxy sample, which has a redshift resolution $\delta z$. Thus, we can compute:
\begin{equation}\label{Nmu}
N_{\mu} = \left\lceil\frac{\Delta z}{\delta z}\right\rceil
\end{equation}
where $\Delta z = z_2 - z_1$ and $z_2$ and $z_1$ are, respectively, the upper and lower limits of the redshift range spanned by our galaxy sample. We select, at each $N_{\mu}$ location, $N_{\sigma}$ values for the standard deviation that linearly span the range from a minimum value of $\sigma_{\rm min}$ to a maximum value $\sigma_{\rm max}$. The minimum value is  selected in such a way that we will approximately have a single Gaussian that fills a single redshift bin of width $\delta z$. In practice, a Gaussian vanishes at approximately 3$\sigma$ from the mean; therefore, we can select $\sigma_1 = \delta z/6$. 

On the other hand, we select the broadest basis function to approximately cover half of the full redshift range $\Delta z$ at each position; therefore, we select $\sigma_{\rm max}  = \Delta z/12$. Although the extreme basis functions are not  frequently used, they ensure that we cover all possibilities. Finally, we set the resolution between different values of $\sigma$ to be $\delta z /2$ in order to make sure the difference between two consecutive Gaussian basis functions is on the order of 
$\delta z$. Setting $\Delta \sigma = \sigma_{\rm max} - \sigma_{\rm min}$  we have that $N_{\sigma}$ is given by:
\begin{equation}\label{Nsigma1}
N_{\sigma} = \left\lceil\frac{2 \Delta \sigma}{\delta z}\right\rceil
\end{equation}
which can be simplified to
\begin{equation}\label{Nsigma2}
N_{\sigma} = \left\lceil \frac{\Delta z}{6 \delta z} - \frac{1}{3}\right\rceil \approx \frac{N_{\mu}}{6}
\end{equation}

As some \pz PDFs have extended wings, we also generate $N_{\gamma}$ basis functions for each Gaussian basis function with extended profiles by using a Voigt profile. Voigt profiles are widely used in spectral line fitting, and are defined as the convolution between a Gaussian distribution and a Lorentzian distribution. A Voigt profile can be written as the real part of the Faddeeva function~\citep{Abramowitz1972}:
\begin{equation}
 V(x;\sigma,\gamma) = \frac{1}{\sigma\sqrt{2\pi}} \operatorname{Re}\left[e^{-z^2} \left(1-\erf (-iz)\right)\right]
\end{equation}
where $\erf(-iz)$ is the complex \textit{error function}. $z=\frac{(x-\mu)+i\gamma}{\sigma\sqrt{2}}$ is a complex variable, where $\mu$ is the center of the function, $\sigma$ is the standard deviation from the Gaussian, and $\gamma$ determines the strength of the extended wings and is a parameter from the Lorentz distribution. As a result, if $\gamma = 0$, we have a Gaussian distribution with parameters $\mu$ and $\sigma$. 

We present examples of different Voigt profiles in Figure~\ref{fig:voigt} given a fixed $\mu= 0.3$ and $\sigma=0.01$, but with $\gamma$ varying from zero (Gaussian) to one $\sigma$.  We do not, however, select pure Lorentzian profiles, as they produce distributions that are too extended to be practical for this analysis. In practice, we find that an upper limit of $\gamma = 0.5\sigma$ is sufficient to accurately model any extended wings. Thus, including the Gaussian case with $\gamma=0$, we fix $N_{\gamma} =6$ and allow $\gamma$ to vary linearly from $0$ to $0.5 \sigma$ in steps of $0.1\sigma$. Thus, in the most simple case we would only consider basis functions with $\gamma = 0$ and $N_{\gamma} =1 $.

\begin{figure}
\includegraphics[width=0.48\textwidth]{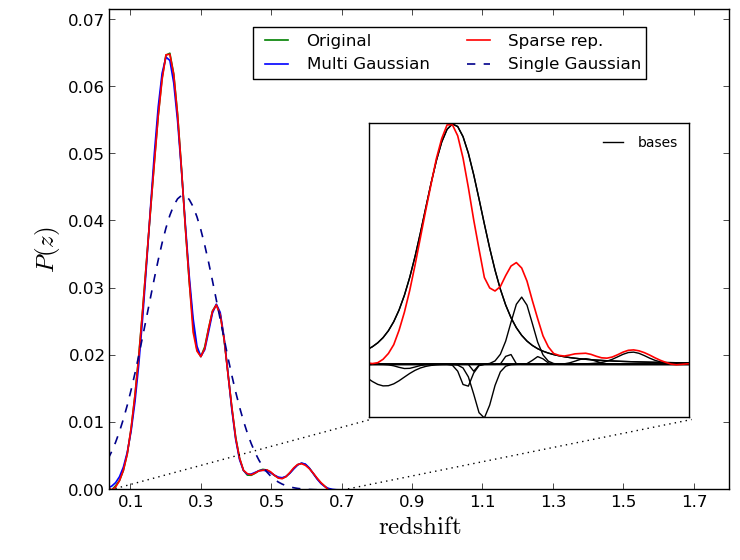}
\caption{The representation of an original \pz PDF (green) given by three techniques: multi-Gaussian (blue), single Gaussian (blue dashed line), and sparse basis representation (red). The inset panel shows the final bases (in black) used to represent the \pz PDF while the recovered distribution is shown in red.}
\label{fig:example}
\end{figure}

In total, the dictionary is composed of $ N_{\rm total} = N_{\mu} \times N_{\sigma} \times N_{\gamma}$ bases, which all have $\ell_2$ norm equal to unity. By using our previous definitions, we have the following approximate rule of thumb for creating a dictionary:
\begin{equation}
N_{\rm total} \approx N_{\mu}^2 = \left( \frac{\Delta z}{\delta z}\right)^2
\label{bfapprox}
\end{equation}
Although this is an estimate, it provides a very good approximation to the total number of bases needed given the resolution of the original \pz PDF. Additional bases are not necessary and little is gained by using a finer resolution. \PZ codes generally provide \pz PDFs by using roughly two to three hundred points. According to Equation~\ref{bfapprox}, we notice that for 250 sample points in a PDF, we would need approximately 62,500 bases. Thus, we can use a 2-byte integer to express the indices into our basis function dictionary, which has important ramifications in the compact storage of \pz PDFs as discussed in the next section.

\section{Discussion}

We have applied the previously discussed \pz PDF representation techniques to the CFHTLenS data introduced in Section~\ref{data}. We have computed a \pz PDF for each galaxy in the one million test sample by using the \tpz software to compute a PDF with two hundred sampled points at a resolution of $\delta z = 0.011$. We display one such \pz PDF in Figure \ref{fig:example} where the original distribution is shown in green, a multi-Gaussian representation is shown in blue, a sparse basis representation is shown in red, and a single Gaussian model is shown with a blue dashed line. We can see that both the sparse basis representation and the multi-Gaussian agree remarkably well with the original \pz PDF, to the point where it is hard to see the original PDF. As one would expect in this multi-peak PDF, the single Gaussian model does not reproduce this \pz PDF very well. The inset panel provides a zoomed-in view showing the sparse basis representation of the \pz PDF and the actual basis functions used in the 
representation. As the number of bases is increased, we expect some of them to have a negative coefficient, as shown in the inset, which aids in the reconstruction of the residuals from the previous bases. Given the iterative nature of this process, we select the new basis function that optimally corrects the residuals of previous bases in order to best reconstruct the \pz PDF by using the minimum number of functions.

\begin{figure}
\includegraphics[width=0.48\textwidth]{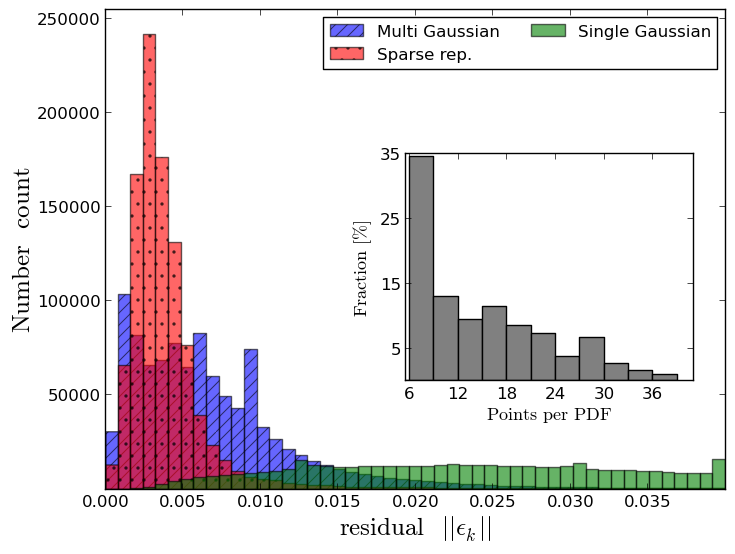}
\caption{The residual distribution for all CFHTLenS galaxies computed by using the multi-Gaussian representation (blue) and the sparse basis representation (red). In each case, we use the same number of representation values for each galaxy \pz PDF. For comparison, the Single-Gaussian representation is shown in green. \textit{Inset}: The distribution of points (bases or fitting parameters) per galaxy \pz PDF. The number of peaks, $Np_k$, per \pz PDF is the same, but divided by 6. Thus, there are $3(Np_k+1)$ parameters per galaxy.}
\label{fig:same_number}
\end{figure}

In order to quantitatively compare the reconstruction of the \pz PDF by using the three methods as shown in Figure \ref{fig:example}, we compute the multi-Gaussian fitting for all $10^6$ galaxies from our CFHTLenS test sample. For each galaxy we record the number of values (parameters) required to accurately reconstruct the original PDF. Note that in this fitting approach, we are not fixing the number of Gaussian functions used in the reconstruction, but are instead defining the number of Gaussians as the number of peaks in the \pz PDF plus one extra Gaussian to compensate for the residuals and extended profiles. In addition, we also compute, for each galaxy, the optimal sparse representation by using a variable number of basis functions that are constrained to match the number of points used in the multi-Gaussian fitting. We also compute the best single Gaussian fit to each PDF to demonstrate the importance in using the information contained within the full PDF as opposed to simply treating each \pz 
estimate as a Gaussian PDF.

After computing the different representations for each galaxy, we next compute the norm of the residuals between each representation and the original \pz PDF for each galaxy and accumulate the results. We compare the resulting distributions in Figure~\ref{fig:same_number}. First, we notice the broad shape of the single Gaussian distribution (green). In fact, the width of the distribution exceeds the plot boundaries as the median of the single Gaussian distribution of residuals is 0.043, which is outside the range of the Figure. Second, we observe that when using the same number of values to represent the \pz PDF, the sparse basis representation produces much smaller residuals with a more concentrated distribution than the multi-Gaussian fitting. Specifically, the median of the sparse representation residual distribution is 0.0033, which is almost half of the value (0.0058) for the multi-Gaussian fitting. Both of these results indicate that either method provides a good representation of the \pz PDF by using 
a small number of values. We also show the distribution of values required to reconstruct the \pz PDF of each galaxy in the inset panel of Figure~\ref{fig:same_number}. This subplot indicates that approximately 35\% of the galaxies are single peaked (six values are required for two Gaussians, which in our implementation means a single peak plus an extra Gaussian for the extended wings). The distribution extends up to thirty-nine values for roughly 1-2\% of the sample, which corresponds to twelve peaks in a \pz PDF. The average number of values per galaxy is fourteen, which, in itself, implies a large compression ratio when compared to the original two hundred values while still providing a very good reconstruction of the full \pz PDF.

\begin{figure}
\includegraphics[width=0.48\textwidth]{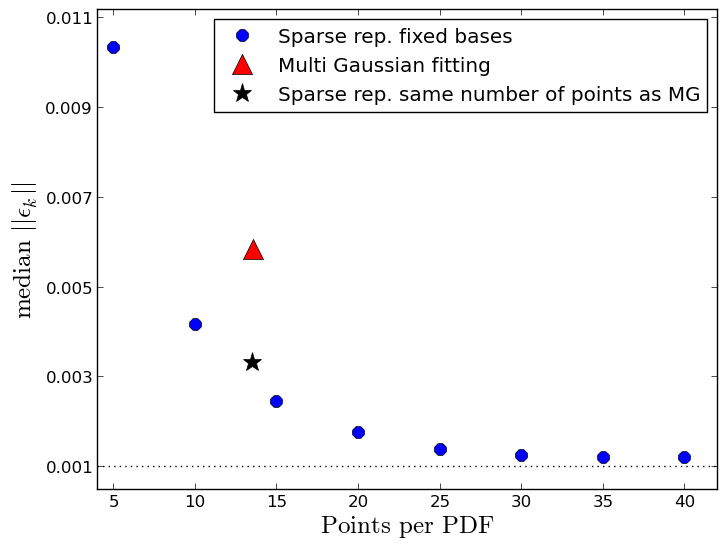}
\caption{The median of the residual distribution as a function of the number of fixed bases used to reconstruct each galaxy's \pz PDF when using the sparse representation technique (blue dots). For reference, the median of the multi-Gaussian residual distribution (red triangle) and the median of the sparse representation with variable number of bases (black star) are also shown, where on average both techniques need fourteen points per \pz PDF.}
\label{fig:sparse_summary}
\end{figure}

While a natural number of basis functions can be determined for the multi-Gaussian representation, the sparse basis representation is more general and thus does not have a simple, natural number of basis functions. In order to better understand the optimal number of basis functions for \pz PDFs, we compute the sparse basis representation for all galaxies in the test sample by using a different number of fixed bases. We combine the residuals, and plot the median value of the distribution as a function of the number of values used to represent the PDF as blue dots in Figure \ref{fig:sparse_summary}. As shown in the figure, as we increase the number of bases, the residuals decrease monotonically. This decrease is quite rapid at first, as expected, and slowly decreases until approximately twenty-five bases are used. For comparison, we also show the multi-Gaussian residuals for fourteen values (red triangle) and the corresponding sparse basis representation residuals for approximately the same number of values (
black 
star), demonstrating the superiority, in terms of precision, of the sparse representation over the multi-Gaussian. If we restrict the number of values to twenty, we have a median residual of 0.018, which corresponds to a median reconstruction of all one million test galaxies at 99.82\% at a resolution of $\delta z = 0.011$. Since the original \pz PDF contained two hundred points, this implies a compression ratio of ten.

\begin{figure}
\includegraphics[width=0.48\textwidth]{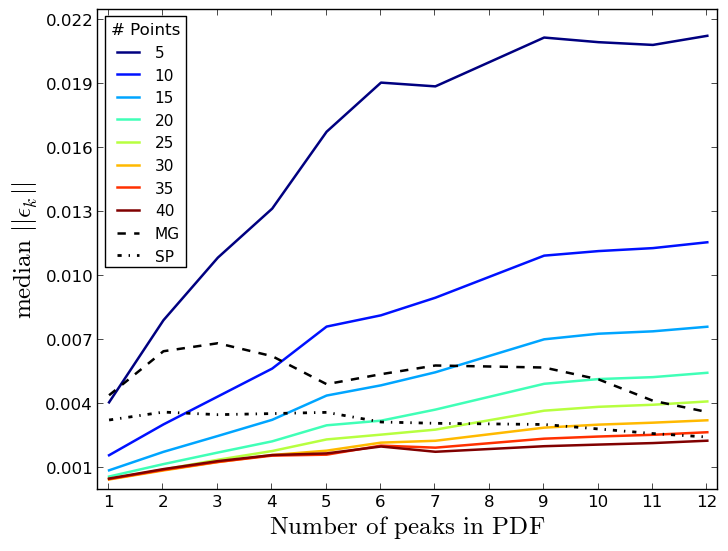}
\caption{The median of the residual distribution as a function of the number of peaks in the \pz PDF when using (solid color lines) a different number of fixed bases in the sparse basis representation, (black dashed line) when using the multi-Gaussian fitting technique, and (black dashed-dotted line) when using the sparse representation when the number of bases is equivalent to the number of multi-Gaussians.}
\label{fig:res_peak}
\end{figure}

Clearly these results will vary depending on the galaxy sample. In particular, the data we use in this analysis are from the CFHTLenS, which is a representative deep survey with galaxies that have \pz PDFs with up to twelve peaks. The performance of the sparse representation also depends directly on the number of peaks in each PDF when we globally fix the number of bases. In Figure~\ref{fig:res_peak}, we display the median of the residual distribution as a function of the number of peaks in the \pz PDF, with different curves corresponding to different numbers of globally fixed bases. For a fixed number of bases, the residual increases as the number of peaks increase. Thus, a galaxy sample that consistently has a low number of peaks will have increased performance when using a smaller number of bases. 

For example, we  achieve a 99.5\% reconstruction by using only ten values for galaxies with four or fewer peaks. In \cite{CarrascoKind2014b}, we discussed the relationship between the number of peaks and the shape of the \pz PDFs with the outlier fraction. With this in mind, we could reduce the number of bases used to reconstruct a sample and flag those with a high number of peaks, where the reconstruction is less reliable, for further investigation. In fact, we achieve a reconstruction of 99\% for \pz PDFs with three or fewer peaks when using only five bases for the sparse representation. This produces a compression ratio of forty when the original \pz PDF has two hundred points.

For comparison, we also show the fitting residuals for the multi-Gaussian (black dashed line) and sparse representation (black dashed-dotted lines) where the variable number of bases matches the number of multi-Gaussians. The performance of the multi-Gaussian fitting is less dependent on the number of peaks simply because the number of parameters dynamically changes for each \pz PDF. Overall, the multi-Gaussian performance is fairly consistent at around 0.005, even as we increase the number of peaks. The sparse representation with a variable number of bases, on the other hand, is less dependent on the number of peaks and has residuals that are nearly 50\% smaller than the multi-Gaussian fitting at an approximately constant value of 0.003.

\subsection*{PDF Storage}

In the previous section, we discussed how the sparse representation and the multi-Gaussian fitting can accurately represent a \pz PDF by using only a few dozen values with a reconstruction level of 99\%. In the case of the multi-Gaussian fitting, the number of parameters to be stored will depend on the number of peaks in each individual PDF. As discussed previously, we will have $3(Np_k+1)$ parameters, which are all floating point numbers. For this dataset we found that the average number of values (or floating point parameters) required is fourteen; but to store these data for all galaxies, we would need to combine the results from different galaxies in order to take advantage of the galaxies that require fewer values so that we can also store those galaxies that require a larger number of parameters. Varying the number of values to store galaxy \pz PDFs in this manner might not be practical, as it will likely depend strongly on the archival and storage system while also increasing the computational 
difficulty in dealing with a varying number of parameters for different \pz PDFs. The practical solution would be to use thirty-nine fixed values (the maximum required for this dataset) for all galaxies and store them independently. This result is also true for the varying sparse representation, which we have demonstrated has a better performance in comparison to the multi-Gaussian when representing a \pz PDF.

On the other hand, requiring a fixed number of basis functions per galaxy alleviates this issue and also has the additional benefit that there is no need to pad with zeros since having more points for single peaked galaxies simply provides a more accurate representation. We have shown that by using ten to twenty values we are able to produce a residual on the order of 0.1\%, where all galaxies are stored independently. One additional (and very important) advantage of the sparse basis representation is that all bases in the dictionary have $\ell_2$ norm equal to unity. Furthermore, when bases are computed by using the OMP algorithm, the absolute values of all coefficients are, by definition, less than unity. They can be negative, however, as seen in Figure \ref{fig:example}. Since the PDFs are probability distributions, by definition the integral of the PDFs over the redshift range must also be unity. As a result, we can rescale all coefficients; and, as long as their relative amplitudes are the same, we can 
always impose the integral normalization at the end of the reconstruction. 

If we continue this line of reasoning, we can rescale the coefficients of every basis function for a given galaxy so that the coefficients have absolute values between zero and one. When doing this  we will be sure that the first basis function has unit amplitude without loss of accuracy on the very first basis. We can discretize this range by using approximately 32,000 sampling points (specifically $2^{15}$) between zero and one, and store the corresponding integer from this range, and its sign, in a single sixteen-bit value. The error introduce by this discretization is very small, on the order of $10^{-5}$, which is almost always negligible for most applications. In this approach, the most important basis is always first, and since it defines the scale, is always stored with no rounding errors. 

\begin{figure}
\includegraphics[width=0.48\textwidth]{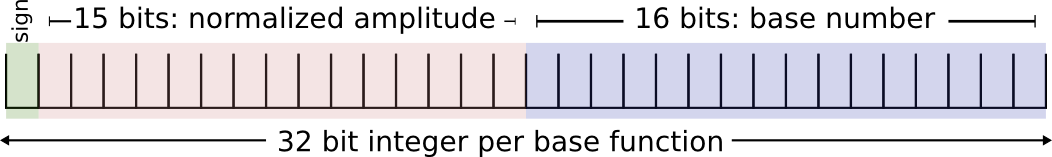}
\caption{A single four-byte integer scheme to store a single basis function in the sparse representation method. The first sixteen bits store the coefficients (including sign), while the second sixteen bits store the location of the bases in the dictionary.}
\label{fig:32bit}
\end{figure}

We have, in fact, used this discretization throughout this paper; the difference introduced by using this discretization and the real values is less than 0.0005\% and thus it does not directly affect the representation accuracy. This allows us to only use one value per basis function in our sparse representation. Since our dictionary contains fewer than 65,000 bases, which can be completely represented by a sixteen-bit integer, we can use a single four-byte integer, as shown graphically in Figure~\ref{fig:32bit}, to store both the base function and its amplitude and sign. 
More specifically, if we have a two hundred point \pz PDF, which corresponds to a resolution of $\delta z = 0.011$ over the range $z=0$ to $z=2.2$, and we fix our representation to use ten bases, we can achieve an average reconstruction accuracy of 99.5\% by using only 40Bs per \pz PDF. Given a million galaxies that we treat in this manner, we will only need approximately 38MBs to store all of their PDFs. In addition, since we are only storing four-byte integers to represent the full \pz PDF, we can potentially reduce the overall disk storage requirements by employing existing bit compression techniques~\citep{Lemire2012} which will be important for relational database systems.

The representation and data encoding scheme we have proposed is, of course, even more flexible than we have demonstrated. If our \pz PDFs employ either a different redshift resolution, span a larger redshift range, or simply have been sampled at a higher number of points, we can still use a four-byte integer representation. For example, if the original PDF is sampled at a finer resolution, we can double $N_{\mu}$ and reduce $N_{\gamma}$ by one-half and still retain the same number of bases, recall we simply need the number of bases to be less than $2^{16}$ (or 65,536) in order to still have $2^{15}$ bits to encode the basis function index. In an extreme case, we can revert to a purely Gaussian set of basis functions and allow $N_{\mu}$ and $N_{\sigma}$ to vary while keeping the total number of bases below the $2^{16}$ limit. In this case, we likely would need to increase the number of fixed bases in order to accurately represent the \pz PDF.

If the number of required bases exceeds the $2^{16}$ limit, because, for instance, our \pz PDFs are sampled at an extremely high resolution or span a large redshift range, we can always increase the size of the dictionary beyond this two-byte limit. In this case, we simply have a very dense dictionary, where fewer fixed bases would be necessary; thus, each basis function would be stored in either a six-byte or an eight-byte integer, depending on the details of the computational system. Another alternative would be to fix the number of bits used to encode each type of basis function; for example, to use two-bits for $N_{\gamma}$, six-bits for $N_{\sigma}$ and eight-bits for $N_{\mu}$, resulting in four, sixty-four, and two hundred and fifty-six possible values for each basis function. As a fixed framework, this technique could also simplify the storage and functional indexing. Finally, as we have mentioned earlier, there is no need to store the entire dictionary since it is simply defined over a functional 
basis. Instead, we only need to store the parameters required to regenerate the dictionary so that we can either regenerate the dictionary or generate the individual functions themselves as needed.

\section{Application}

We now change our focus from the computation of a compact representation of a \pz PDF to a demonstration of their use in a scientific analysis. Perhaps the simplest application of a \pz PDF is the computation of the number of galaxies as a function of redshift, which is widely used in the statistical analysis of the spatial distribution of galaxies. This function is computed by binning spectroscopic observations of galaxies as a function of redshift; but for a photometric survey, this distribution is optimally computed by integrating over all individual \pz PDFs at a given resolution. This approach, while more computationally demanding, has been shown to be better than simply using a single estimate for the \pz~\citep[\eg][]{Mandelbaum2008,Sheldon2012,CarrascoKind2013}. Even with this simple application, however, we benefit from the use of a sparse representation for our \pz PDFs, since we can transform our theoretical framework to use our basis functions. Thus we can operate directly over the dictionary 
and use the sparse basis indices and coefficient parameters to calculate the true and reconstructed values taking into account the normalization.

\begin{figure*}
\includegraphics[width=0.98\textwidth]{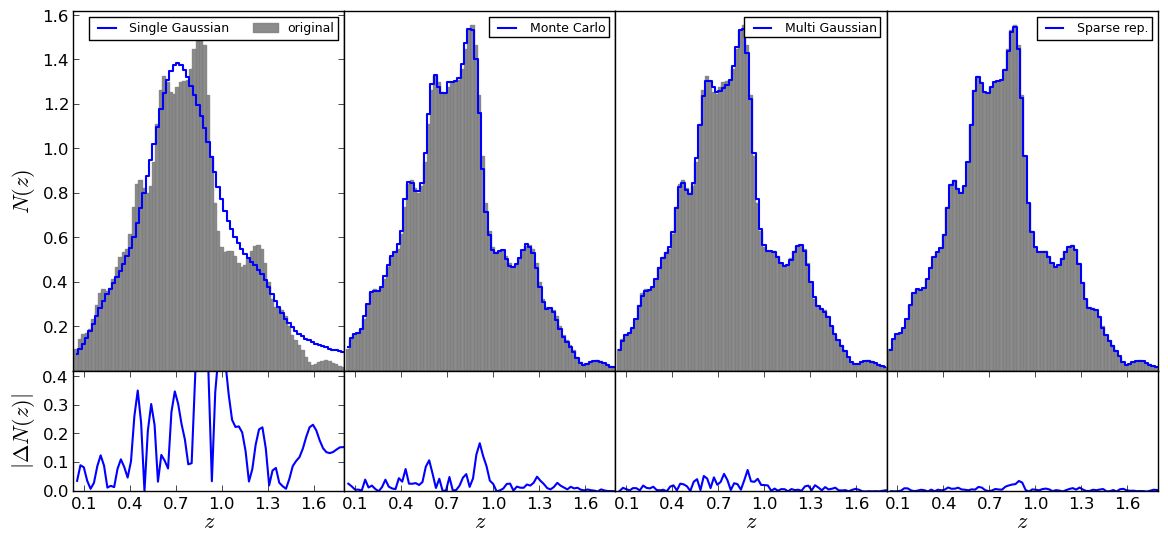}
\caption{The $N(z)$ distribution for all $10^6$ galaxies from the CFHTLenS data computed by using the four \pz PDF representation techniques. Within each panel, the original $N(z)$ computed by stacking the full \pz PDFs is shown in gray and a different representation method is shown in  blue. From left to right we have the single Gaussian model, the Monte Carlo sampling described in Section \ref{singlez}, the multi-Gaussian fitting method, and the sparse representation method that uses the same number of bases as the multi-Gaussian method. The bottom panels show the absolute difference between the original and reproduction at the same scale.}
\label{fig:diff_nz}
\end{figure*}

As a demonstration, we can compute the galaxy redshift distribution directly over the basis functions. We start by writing the definition of $N(z)$:
\begin{equation}\label{AP1}
N(z) = \sum\limits_{k=1}^{N}\int_{z-\Delta z/2}^{z+\Delta z/2} P_k(z)dz
\end{equation}
where the sum is over all $N$  galaxies and $P_k(z)$ is the \pz PDF of a given galaxy $k$. $z$ is the midpoint of each redshift bin, which have a fixed width $\Delta z$. We can rewrite this equation in terms of the PDF representation for $P_k(z)$, which we previously defined as $\vec{pz}_k$. Thus, in the sparse basis representation, we can express each PDF as:
\begin{equation}
\vec{pz}_k \approx \vec{D} \cdot \boldsymbol{\delta}_k
\end{equation}
where $\boldsymbol{\delta}_k$ is a sparse vector, which might contain ten to twenty elements, that contains the amplitudes for each functional basis and $\vec{D}$ is an $n \times m$ dictionary, where $n$ is the number of points in the original PDF and $m$ is the total number of bases. By using this result, we can rewrite Equation \ref{AP1}:
\begin{equation}\label{AP2}
N(z) = \sum\limits_{k=1}^{N} \boldsymbol{\delta}_k  \cdot \int_{z-\Delta z/2}^{z+\Delta z/2} \vec{D} dz
\end{equation}
where $\boldsymbol{\delta}_k$ is independent of redshift so that we only need to integrate once over each basis function in the dictionary; thus, we only have $m$ integrations instead of $N$.

Furthermore, we can precompute this integral over $\vec{D}$, which we denote by $\vec{I_D}(z)$. This integral corresponds to a vector of length $m$, where each entry is the integral over each one of the basis functions $\vec{d}_j$ in $\vec{D}$:
\begin{equation}
\vec{I_D}(z) = \int_{z-\Delta z/2}^{z+\Delta z/2} \vec{d}_j dz \qquad {j=1,2,\dots, m}
\label{eq:dbint}
\end{equation}
Since all $N$ galaxies are expressed in terms of the $m$ bases, we can also pre-factorize the coefficients (or amplitudes) in a vector $\boldsymbol{\delta}_N$:
\begin{equation}
 \boldsymbol{\delta}_N = \sum\limits_{k=1}^{N} \boldsymbol{\delta}_k
\end{equation}
Therefore, after these precomputations we can simply express $N(z)$ as:
\begin{equation}\label{lastAP}
N(z) =  \vec{I_D}(z) \cdot \boldsymbol{\delta}_N 
\end{equation}
reducing the computation to a simple dot product of precomputed quantities. For each bin, we need to compute $\vec{I_D}(z)$, but $\boldsymbol{\delta}_N$ is computed only once and can be used both for all bins in the computation of $N(z)$ and in other cosmological applications. This result is also true for other linear operations that might be involved in another cosmological analysis. Thus, by working directly in the space defined by the basis functions, we can reduce computational memory and processing times significantly.

We compare the original $N(z)$ for $10^6$ test galaxies from the CFHTLenS sample to different $N(z)$ distributions reconstructed by using Equation~\ref{lastAP} for different representation formats in Figure \ref{fig:diff_nz}. Each original galaxy \pz PDF has a resolution of $\delta z = 0.011$ and contains two hundred values. We restrict the comparison in Figure \ref{fig:diff_nz} to four techniques: a single Gaussian model, the Monte Carlo estimator described in Section~\ref{singlez}, a multi-Gaussian fitting technique, and the sparse basis representation. However, we compute the fractional percentile error between the original $N(z)$ and all eight techniques and compare the results in Table~\ref{tab:Nz_err}. Before discussing the performance of individual techniques, we note that the lower panels in Figure~\ref{fig:diff_nz} are all shown at the same scale to facilitate direct comparisons.

In the first panel, we see that the single Gaussian model clearly shows a significant difference, which is visible both from the distribution itself and in the bottom panel from the absolute error between these two distributions. Next, we see that the single point \pz estimation computed by using a Monte Carlo sampling shows a surprisingly good agreement with the original distribution. This result was discussed by \cite{Wittman2009}, who demonstrated that this technique does provide a fair statistical representation of the sample's galaxy redshift distribution. This approach, where the $N(z)$ distribution is computed as a random sample drawn from the cumulative PDF of each galaxy, statistically compensates for the \pz errors for an individual galaxy and thus produces a reliable $N(z)$ distribution. This approach does, however, introduce much larger errors on the estimation of individual galaxy \pzsns. While one might be tempted to store a \pz PDF by using this approach in order to accurately recover an $N(z)
$ by storing a minimum quantity of new data, it would be easier to simply compute and store the actual $N(z)$. Furthermore, since this technique is dependent on using a large number of galaxies to generate a more accurate $N(z)$, if one is interested in computing the redshift distribution for galaxy subsets, the reconstruction accuracy might suffer as the number of galaxies in the subsamples is decreased.


In the third panel, we see that the multi-Gaussian fitting technique has a smaller error than the first two methods. As discussed previously, this method provides an accurate representation of a \pz PDF, thus it would be expected to also yield an accurate representation for $N(z)$. Finally, in the last panel we have the results for the sparse basis representation where the number of bases used is defined to be the same as required for the multi-Gaussian fitting method. As seen previously with the distribution of residuals, we see that, with this direct cosmological application, we recover the original $N(z)$, by using the same number of values to represent the \pz PDF, more accurately than with other techniques.

\begin{figure}
\includegraphics[width=0.48\textwidth]{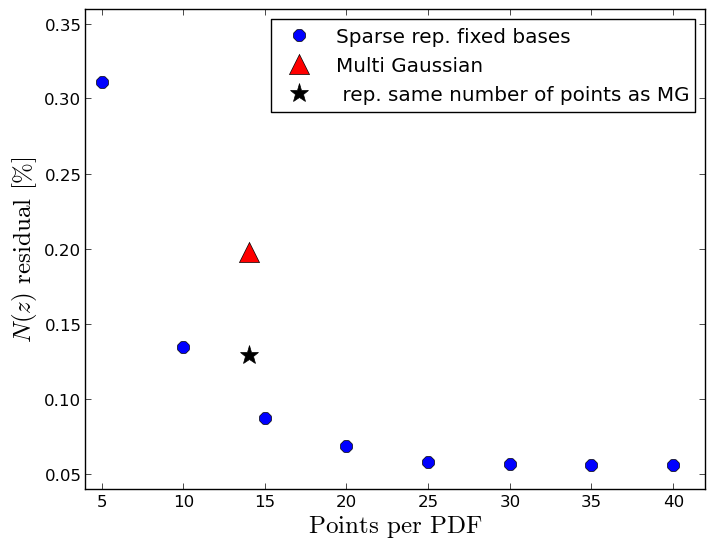}
\caption{The fractional percentile error between the original $N(z)$ and the reconstructed $N(z)$ computed by using Equation~\ref{lastAP} when fixing the number of bases used to represent the \pz PDFs for all galaxies with the sparse representation technique (blue dots). For comparison, we also show the multi-Gaussian (red triangle) and the sparse representation with a variable number of bases (black star) residuals. On average, both of these latter techniques require fourteen points per \pz PDF.}
\label{fig:NZ_sparse}
\end{figure}

We present the fractional percentile error between the original $N(z)$ and the reconstructed $N(z)$ computed by using Equation~\ref{lastAP} for different \pz PDF representation techniques in Figure~\ref{fig:NZ_sparse}. As also seen in Figure~\ref{fig:sparse_summary}, we see that as the number of bases increases for the sparse representation (shown in blue dots) the accuracy of the reconstruction also improves, but here we focus on the error in the reconstruction of $N(z)$. Since additional bases will produce a more accurate \pz PDF representation, we also expect a more accurate $N(z)$ reconstruction when the number of bases increases. For comparison, we also show the multi-Gaussian fitting (red triangle) and the sparse representation (black star) where the number of bases matches the multi-Gaussian fitting value.  

We observe that, by using only fifteen values, we can reconstruct the $N(z)$ distribution to an accuracy of 99.9\% as measured with respect to the original distribution. In addition, this result changes only slightly when we limit our representation to ten bases. We also see that the error values are slightly better than we saw when reconstructing the individual \pz PDFs, because computing the $N(z)$ smooths over the individual \pz PDFs, thereby reducing the impact from small discrepancies in individual \pz PDFs that might result from using a specific functional basis. If we increase our representation to use forty bases, we can reconstruct the $N(z)$ distribution to nearly 99.95\%, but the decrease in the error, however, does not change significantly once we have used approximately twenty-five bases, suggesting there are diminishing returns.

In Section \ref{singlez}, we introduced several different individual \pz estimates that are widely used, including the mean, the mode, and the median of the \pz PDF, and Monte Carlo sampling from the cumulative \pz PDF. These single estimates show an even larger fractional error than visible on the vertical axis shown in Figure~\ref{fig:NZ_sparse}, and are thus presented in Table~\ref{tab:Nz_err}, which summarize the results from all of the methods presented herein, including the number of values required by the representation method and the fractional percentile error for that method in reconstructing the original $N(z)$. 

The entries in Table~\ref{tab:Nz_err} are presented in ascending order by the size of this fractional percentile error. From these entries, we see that the single Gaussian model has, on average, a reconstruction error of 2.2\% while the single value estimates all have reconstruction errors over 6\%. The Monte Carlo sampling method provides the best reconstruction results when using a single \pz estimate with an error of about 0.4\%, which is comparable  to a sparse representation that uses five bases. As mentioned previously, however, this technique does not provide accurate individual \pz estimates. We also observe that the difference when using thirty, thirty-five, or even forty bases is very small, although it is bigger than the resolution in the discretization scheme; thus our proposed discretization method does not impact these results and we can safely represent each basis function by using a single four-byte integer.

\begin{table}
\caption{The fractional percentile error between the original $N(z)$ and a reconstructed $N(z)$ computed by using the sparse basis representation, single and multi-Gaussian fitting, and all of the single point \pz techniques described in Section \ref{singlez}. We also list the number of values required to represent the \pz PDF. The table is sorted in ascending order by the percentile error.}
\label{tab:Nz_err}
\renewcommand{\footnoterule}{}
\begin{tabular}{lcl}
Method & Values per PDF  & Error [\%] \\
\hline
sparse rep. fixed&  40 & 0.05545 \\
sparse rep. fixed&  35 & 0.05551 \\
sparse rep. fixed&  30 & 0.05611 \\
sparse rep. fixed&  25 & 0.05750 \\
sparse rep. fixed&  20 & 0.06829 \\
sparse rep. fixed&  15 & 0.08729 \\
sparse rep. same MG&  14 & 0.12930 \\
sparse rep. fixed&  10 & 0.13440 \\
multi-Gaussian&  14 & 0.19779 \\
sparse rep. fixed&  5 & 0.31113 \\
Monte Carlo&  1 & 0.37294 \\
single Gaussian&  2 & 2.19095 \\
Median PDF&  1 & 6.63550 \\
Mean PDF&  1 & 7.47077 \\
Mode PDF&  1 & 13.24271 \\
\hline
\end{tabular}
\end{table}

The integration over the dictionary of bases, as shown in Equation~\ref{eq:dbint}, can also be used to compute $N(z)$ over different redshift bins. In this case, the integration can be performed by using the bases and subsequently applying Equation~\ref{lastAP} when using the sparse basis representation. Furthermore, we can extend this approach to analyze multiple \pz PDFs for each galaxy, where they are each represented by the same dictionary. This would prove useful when a survey has stored \pz PDFs for the same galaxy by using different galaxy spectral templates. Thus, a scientist could either compute an $N(z)$ by using a single, per-galaxy \textit{best} template or compare different $N(z)$ that are computed by using different template combinations. Since the integrals could all be precomputed, the only new computation is for the basis coefficients for each galaxy, dramatically reducing the overall computational demands.

For example, we might have different (or even updated) priors for different galaxy types in a survey. We can quickly apply these new priors to the precomputed dictionary integrals and recover the results for each galaxy given their basis coefficients in an efficient manner. Alternatively, one might want to  minimize over the galaxy type under certain restrictions, which can be applied over the precomputed integrals of the dictionary of bases. The minimization problem subsequently becomes a simple task of selecting the minimum or maximum sum over the coefficients. As should be evident from these examples, there exist a number of different applications where our proposed sparse basis representation not only reduces the overall storage requirements, often significantly, but also reduces the computational requirements for cosmological analyses.

\section{Conclusions}

In this paper, we have presented different techniques to represent and efficiently store \pz PDFs, which have been shown to convey significantly more information than a single \pz estimate. As we enter the era of precision cosmology, the growth of large, dense photometric surveys has created an unmet need to quantify and manage these probabilistic values for hundreds of millions to billions of galaxies. Specifically, we have introduced the use of a sparse basis representation that uses a dictionary of Gaussian functions and Voigt profiles, which have extended wings, to accurately and efficiently represent each \pz PDF. We minimize the number of required bases while maintaining a high accuracy by using an Orthogonal Matching Pursuit algorithm, which provides a unique set of bases for each \pz PDF while minimizing the residual between the original and final \pz PDF. This algorithm is publicly available\footnote{http://lcdm.astro.illinois.edu/code/pdfz.html}

We use photometric data from the CFHTLenS survey to compute \pz PDFs by using our \tpz code, producing PDFs with two hundred points and a redshift resolution of $\delta z = 0.011$. By using these PDFs, we demonstrate the our proposed sparse basis representation reconstructs a more accurate PDF than other techniques, include a multi-Gaussian fitting approach with a flexible number of parameters based on the number of peaks in each PDF. If we use the exact same number of parameters with our sparse representation as used by the multi-Gaussian fitting, we found that the sparse basis representation results are superior with the additional benefit that each basis or parameter can be stored using a single integer. We also showed that, with a fixed number of bases, we could achieve both a highly accurate PDF that also has a large compression ratio. As a specific example, we found that by using only ten (twenty) values per \pz PDF, we could reconstruct a \pz PDF at over a 99.5\% accuracy with a compression ratio of 
twenty (ten), providing a significant storage reduction without a loss of information.

We quantified the number of bases required within the sparse representation dictionary, specifically finding that $(\Delta z/ \delta z)^2$ bases are sufficient to represent the galaxy \pz PDFs in our CFHTLenS test sample, where $\Delta z$ is the overall redshift range and $\delta z$ is the photometric redshift PDF resolution. If the number of points in the original PDF is approximately $200$--$250$, we can use a dictionary with fewer than $2^{16}$ bases, which results in an accurate PDF reconstruction while only requiring a single sixteen-bit integer to store the basis index in the dictionary. Furthermore, since the bases themselves are normalized, all basis coefficients are less than unity by definition; and, since the \pz PDFs are also normalized, we only need to retain the relative amplitudes of each basis function. 

Therefore, we can independently rescale the coefficients for each galaxy to their maximum value and subsequently represent them by using a discretized range containing $2^{15}$ values. This will provide a resolution less than $10^{-5}$, and since we set the maximum value from the most significant basis function, it is always correctly represented. As a result, we can also store the coefficients (sign included) in a separate sixteen-bit integer without losing information. Taken together, we can completely encode a single basis function, both dictionary index and coefficient, in a single four-byte integer, simplifying the data management and significantly reducing the data storage and reconstruction computational requirements.

Of course the results we have presented will depend on the quality of the \pz PDFs to which they are applied, which themselves depend on the details of the \pz algorithm that generated them. As would naively be expected, single peaked \pz PDFs are most accurately reconstructed by using either a multi-Gaussian fitting or a sparse basis representation, where only five points per \pz PDF is sufficient to achieve a 99\% accurate reconstruction. Furthermore, we also demonstrated that, as a simple cosmological application of our \pz PDF reconstruction, we could accurately recover the underlying $N(z)$ distribution. In particular, we recovered the $N(z)$ of our CFHTLenS test sample to an accuracy of 99.87\% by using only ten points per \pz PDF. 

Given their compact nature and the fact that they are predetermined, we showed that we could obtain the sample $N(z)$ by integrating the bases over the sample redshift range and later multiplying by the basis coefficients, which can also be prefactored, thereby significantly reducing the number of required integrations. This same principle can be applied to other linear combinations of \pz PDFs or to more complex analyses if they can be expressed in terms of the underlying bases. This later topic is the subject of future work. Overall, these results are very promising, as current and future photometric surveys will produce up to tens of billions of \pz PDFs. Our proposed approach will either allow a reduction in the overall storage requirements or increase the number of \pz PDFs that can be persistently maintained for each galaxy without increasing the required amount of storage. As a result, this new approach will enable science that would have otherwise been difficult or impossible to accomplish.

\section*{Acknowledgements}
The authors thank the referee for a prompt and careful reading of the manuscript and for comments that improved this work. RJB and MCK acknowledge support from the National Science Foundation Grant No. AST-1313415. MCK has been supported by the Computational Science and Engineering (CSE) fellowship at the University of Illinois at Urbana-Champaign. RJB has been supported in part by the Institute for Advanced Computing Applications and Technologies faculty fellowship at the University of Illinois.

This work also used resources from the Extreme Science and Engineering Discovery Environment (XSEDE), which is supported by National Science Foundation grant number OCI-1053575.

This work is based on observations obtained with MegaPrime/MegaCam, a joint project of CFHT and CEA/IRFU, at the Canada-France-Hawaii Telescope (CFHT) which is operated by the National Research Council (NRC) of Canada, the Institut National des Sciences de l'Univers of the Centre National de la Recherche Scientifique (CNRS) of France, and the University of Hawaii. This research used the facilities of the Canadian Astronomy Data Centre operated by the National Research Council of Canada with the support of the Canadian Space Agency. CFHTLenS data processing was made possible thanks to significant computing support from the NSERC Research Tools and Instruments grant program.
\bibliographystyle{mn2e}

\bibliography{pdf_storage}

\begin{thebibliography}{36}
\expandafter\ifx\csname natexlab\endcsname\relax\def\natexlab#1{#1}\fi

\bibitem[{{Abdalla} {et~al}\mbox{.}(2011){Abdalla}, {Banerji}, {Lahav}, \&
  {Rashkov}}]{Abdalla2011}
{Abdalla} F.~B., {Banerji} M., {Lahav} O., {Rashkov} V., 2011, \mnras, 417,
  1891

\bibitem[{{Abrahamse} {et~al}\mbox{.}(2011){Abrahamse}, {Knox}, {Schmidt},
  {Thorman}, {Tyson}, \& {Zhan}}]{Abrahamse2011}
{Abrahamse} A., {Knox} L., {Schmidt} S., {Thorman} P., {Tyson} J.~A., {Zhan}
  H., 2011, \apj, 734, 36

\bibitem[{Abramowitz \& Stegun(1972)}]{Abramowitz1972}
Abramowitz M., Stegun I., 1972, Handbook of Mathematical Functions: With
  Formulas, Graphs, and Mathematical Tables, Applied mathematics series. Dover
  Publications

\bibitem[{{Ahn} {et~al}\mbox{.}(2013){Ahn}, {Alexandroff}, {Allende Prieto},
  {Anders}, {Anderson}, {Anderton}, {Andrews}, {Aubourg}, {Bailey}, {Bastien},
  \& et~al.}]{Ahn2013}
{Ahn} C.~P. {et~al.}, 2013, ArXiv e-prints : 1307.7735

\bibitem[{{Ben{\'{\i}}tez}(2000)}]{Benitez2000}
{Ben{\'{\i}}tez} N., 2000, \apj, 536, 571

\bibitem[{{Bordoloi} {et~al}\mbox{.}(2010){Bordoloi}, {Lilly}, \&
  {Amara}}]{Bordoloi2010}
{Bordoloi} R., {Lilly} S.~J., {Amara} A., 2010, \mnras, 406, 881

\bibitem[{{Bovy} {et~al}\mbox{.}(2011){Bovy}, {Hennawi}, {Hogg}, {Myers},
  {Kirkpatrick}, {Schlegel}, {Ross}, {Sheldon}, {McGreer}, {Schneider}, \&
  {Weaver}}]{Bovy2011}
{Bovy} J. {et~al.}, 2011, \apj, 729, 141

\bibitem[{{Bovy} {et~al}\mbox{.}(2012){Bovy}, {Myers}, {Hennawi}, {Hogg},
  {McMahon}, {Schiminovich}, {Sheldon}, {Brinkmann}, {Schneider}, \&
  {Weaver}}]{Bovy2012}
{Bovy} J. {et~al.}, 2012, \apj, 749, 41

\bibitem[{{Breiman}(2001)}]{Breiman2001}
{Breiman} L., 2001, Machine Learning, 45, 5

\bibitem[{{Carnero} {et~al}\mbox{.}(2012){Carnero}, {S{\'a}nchez}, {Crocce},
  {Cabr{\'e}}, \& {Gazta{\~n}aga}}]{Carnero2012}
{Carnero} A., {S{\'a}nchez} E., {Crocce} M., {Cabr{\'e}} A., {Gazta{\~n}aga}
  E., 2012, \mnras, 419, 1689

\bibitem[{{Carrasco Kind} \& {Brunner}(2013)}]{CarrascoKind2013}
{Carrasco Kind} M., {Brunner} R.~J., 2013, \mnras, 432, 1483

\bibitem[{{Carrasco Kind} \& {Brunner}(2014)}]{CarrascoKind2014a}
{Carrasco Kind} M., {Brunner} R.~J., 2014, \mnras, 438, 3409

\bibitem[{{Carrasco Kind} \& {Brunner}(2014b)}]{CarrascoKind2014b}
{Carrasco Kind} M., {Brunner} R.~J., 2014b, ArXiv e-prints : 1403.0044

\bibitem[{{Collister} \& {Lahav}(2004)}]{Collister2004}
{Collister} A.~A., {Lahav} O., 2004, \pasp, 116, 345

\bibitem[{{Davis} {et~al}\mbox{.}(2007){Davis}, {Guhathakurta}, {Konidaris},
  {Newman}, {Ashby}, {Biggs}, {Barmby}, {Bundy}, {Chapman}, {Coil},
  {Conselice}, {Cooper}, {Croton}, {Eisenhardt}, {Ellis}, {Faber}, {Fang},
  {Fazio}, {Georgakakis}, {Gerke}, {Goss}, {Gwyn}, {Harker}, {Hopkins},
  {Huang}, {Ivison}, {Kassin}, {Kirby}, {Koekemoer}, {Koo}, {Laird}, {Le
  Floc'h}, {Lin}, {Lotz}, {Marshall}, {Martin}, {Metevier}, {Moustakas},
  {Nandra}, {Noeske}, {Papovich}, {Phillips}, {Rich}, {Rieke}, {Rigopoulou},
  {Salim}, {Schiminovich}, {Simard}, {Smail}, {Small}, {Weiner}, {Willmer},
  {Willner}, {Wilson}, {Wright}, \& {Yan}}]{Davis2007}
{Davis} M. {et~al.}, 2007, \apjl, 660, L1

\bibitem[{{Dawson} {et~al}\mbox{.}(2013){Dawson}, {Schlegel}, {Ahn},
  {Anderson}, {Aubourg}, {Bailey}, {Barkhouser}, {Bautista}, {Beifiori},
  {Berlind}, {Bhardwaj}, {Bizyaev}, {Blake}, {Blanton}, {Blomqvist}, {Bolton},
  {Borde}, {Bovy}, {Brandt}, {Brewington}, {Brinkmann}, {Brown}, {Brownstein},
  {Bundy}, {Busca}, {Carithers}, {Carnero}, {Carr}, {Chen}, {Comparat},
  {Connolly}, {Cope}, {Croft}, {Cuesta}, {da Costa}, {Davenport}, {Delubac},
  {de Putter}, {Dhital}, {Ealet}, {Ebelke}, {Eisenstein}, {Escoffier}, {Fan},
  {Filiz Ak}, {Finley}, {Font-Ribera}, {G{\'e}nova-Santos}, {Gunn}, {Guo},
  {Haggard}, {Hall}, {Hamilton}, {Harris}, {Harris}, {Ho}, {Hogg}, {Holder},
  {Honscheid}, {Huehnerhoff}, {Jordan}, {Jordan}, {Kauffmann}, {Kazin},
  {Kirkby}, {Klaene}, {Kneib}, {Le Goff}, {Lee}, {Long}, {Loomis}, {Lundgren},
  {Lupton}, {Maia}, {Makler}, {Malanushenko}, {Malanushenko}, {Mandelbaum},
  {Manera}, {Maraston}, {Margala}, {Masters}, {McBride}, {McDonald}, {McGreer},
  {McMahon}, {Mena}, {Miralda-Escud{\'e}}, {Montero-Dorta}, {Montesano},
  {Muna}, {Myers}, {Naugle}, {Nichol}, {Noterdaeme}, {Nuza}, {Olmstead},
  {Oravetz}, {Oravetz}, {Owen}, {Padmanabhan}, {Palanque-Delabrouille}, {Pan},
  {Parejko}, {P{\^a}ris}, {Percival}, {P{\'e}rez-Fournon},
  {P{\'e}rez-R{\`a}fols}, {Petitjean}, {Pfaffenberger}, {Pforr}, {Pieri},
  {Prada}, {Price-Whelan}, {Raddick}, {Rebolo}, {Rich}, {Richards}, {Rockosi},
  {Roe}, {Ross}, {Ross}, {Rossi}, {Rubi{\~n}o-Martin}, {Samushia},
  {S{\'a}nchez}, {Sayres}, {Schmidt}, {Schneider}, {Sc{\'o}ccola}, {Seo},
  {Shelden}, {Sheldon}, {Shen}, {Shu}, {Slosar}, {Smee}, {Snedden}, {Stauffer},
  {Steele}, {Strauss}, {Streblyanska}, {Suzuki}, {Swanson}, {Tal}, {Tanaka},
  {Thomas}, {Tinker}, {Tojeiro}, {Tremonti}, {Vargas Maga{\~n}a}, {Verde},
  {Viel}, {Wake}, {Watson}, {Weaver}, {Weinberg}, {Weiner}, {West}, {White},
  {Wood-Vasey}, {Yeche}, {Zehavi}, {Zhao}, \& {Zheng}}]{Dawson2013}
{Dawson} K.~S. {et~al.}, 2013, \aj, 145, 10

\bibitem[{{Erben} {et~al}\mbox{.}(2013){Erben}, {Hildebrandt}, {Miller}, {van
  Waerbeke}, {Heymans}, {Hoekstra}, {Kitching}, {Mellier}, {Benjamin}, {Blake},
  {Bonnett}, {Cordes}, {Coupon}, {Fu}, {Gavazzi}, {Gillis}, {Grocutt}, {Gwyn},
  {Holhjem}, {Hudson}, {Kilbinger}, {Kuijken}, {Milkeraitis}, {Rowe},
  {Schrabback}, {Semboloni}, {Simon}, {Smit}, {Toader}, {Vafaei}, {van Uitert},
  \& {Velander}}]{Erben2013}
{Erben} T. {et~al.}, 2013, \mnras, 433, 2545

\bibitem[{{Garilli} {et~al}\mbox{.}(2014){Garilli}, {Guzzo}, {Scodeggio},
  {Bolzonella}, {Abbas}, {Adami}, {Arnouts}, {Bel}, {Bottini}, {Branchini},
  {Cappi}, {Coupon}, {Cucciati}, {Davidzon}, {De Lucia}, {de la Torre},
  {Franzetti}, {Fritz}, {Fumana}, {Granett}, {Ilbert}, {Iovino}, {Krywult}, {Le
  Brun}, {Le F{\`e}vre}, {Maccagni}, {Ma{\l}ek}, {Marulli}, {McCracken},
  {Paioro}, {Polletta}, {Pollo}, {Schlagenhaufer}, {Tasca}, {Tojeiro},
  {Vergani}, {Zamorani}, {Zanichelli}, {Burden}, {Di Porto}, {Marchetti},
  {Marinoni}, {Mellier}, {Moscardini}, {Nichol}, {Peacock}, {Percival},
  {Phleps}, \& {Wolk}}]{Garilli2014}
{Garilli} B. {et~al.}, 2014, \aap, 562, A23

\bibitem[{{Garilli} {et~al}\mbox{.}(2008){Garilli}, {Le F{\`e}vre}, {Guzzo},
  {Maccagni}, {Le Brun}, {de la Torre}, {Meneux}, {Tresse}, {Franzetti},
  {Zamorani}, {Zanichelli}, {Gregorini}, {Vergani}, {Bottini}, {Scaramella},
  {Scodeggio}, {Vettolani}, {Adami}, {Arnouts}, {Bardelli}, {Bolzonella},
  {Cappi}, {Charlot}, {Ciliegi}, {Contini}, {Foucaud}, {Gavignaud}, {Ilbert},
  {Iovino}, {Lamareille}, {McCracken}, {Marano}, {Marinoni}, {Mazure},
  {Merighi}, {Paltani}, {Pell{\`o}}, {Pollo}, {Pozzetti}, {Radovich}, {Zucca},
  {Blaizot}, {Bongiorno}, {Cucciati}, {Mellier}, {Moreau}, \&
  {Paioro}}]{Garilli2008}
{Garilli} B. {et~al.}, 2008, \aap, 486, 683

\bibitem[{{Gwyn}(2012)}]{Gwyn2012}
{Gwyn} S.~D.~J., 2012, \aj, 143, 38

\bibitem[{{Heymans} {et~al}\mbox{.}(2012){Heymans}, {Van Waerbeke}, {Miller},
  {Erben}, {Hildebrandt}, {Hoekstra}, {Kitching}, {Mellier}, {Simon},
  {Bonnett}, {Coupon}, {Fu}, {Harnois D{\'e}raps}, {Hudson}, {Kilbinger},
  {Kuijken}, {Rowe}, {Schrabback}, {Semboloni}, {van Uitert}, {Vafaei}, \&
  {Velander}}]{Heymans2012}
{Heymans} C. {et~al.}, 2012, \mnras, 427, 146

\bibitem[{{Hildebrandt} {et~al}\mbox{.}(2010){Hildebrandt}, {Arnouts}, {Capak},
  {Moustakas}, {Wolf}, {Abdalla}, {Assef}, {Banerji}, {Ben{\'{\i}}tez},
  {Brammer}, {Budav{\'a}ri}, {Carliles}, {Coe}, {Dahlen}, {Feldmann}, {Gerdes},
  {Gillis}, {Ilbert}, {Kotulla}, {Lahav}, {Li}, {Miralles}, {Purger},
  {Schmidt}, \& {Singal}}]{Hildebrandt2010}
{Hildebrandt} H. {et~al.}, 2010, \aap, 523, A31

\bibitem[{{Hildebrandt} {et~al}\mbox{.}(2012){Hildebrandt}, {Erben}, {Kuijken},
  {van Waerbeke}, {Heymans}, {Coupon}, {Benjamin}, {Bonnett}, {Fu}, {Hoekstra},
  {Kitching}, {Mellier}, {Miller}, {Velander}, {Hudson}, {Rowe}, {Schrabback},
  {Semboloni}, \& {Ben{\'{\i}}tez}}]{Hildebrandt2012}
{Hildebrandt} H. {et~al.}, 2012, \mnras, 421, 2355

\bibitem[{{H{\"{o}}gbom}(1974)}]{Hogbom1974}
{H{\"{o}}gbom} J.~A., 1974, \aaps, 15, 417

\bibitem[{{Ilbert} {et~al}\mbox{.}(2006){Ilbert}, {Arnouts}, {McCracken},
  {Bolzonella}, {Bertin}, {Le F{\`e}vre}, {Mellier}, {Zamorani}, {Pell{\`o}},
  {Iovino}, {Tresse}, {Le Brun}, {Bottini}, {Garilli}, {Maccagni}, {Picat},
  {Scaramella}, {Scodeggio}, {Vettolani}, {Zanichelli}, {Adami}, {Bardelli},
  {Cappi}, {Charlot}, {Ciliegi}, {Contini}, {Cucciati}, {Foucaud}, {Franzetti},
  {Gavignaud}, {Guzzo}, {Marano}, {Marinoni}, {Mazure}, {Meneux}, {Merighi},
  {Paltani}, {Pollo}, {Pozzetti}, {Radovich}, {Zucca}, {Bondi}, {Bongiorno},
  {Busarello}, {de La Torre}, {Gregorini}, {Lamareille}, {Mathez}, {Merluzzi},
  {Ripepi}, {Rizzo}, \& {Vergani}}]{Ilbert2006}
{Ilbert} O. {et~al.}, 2006, \aap, 457, 841

\bibitem[{{Jee} {et~al}\mbox{.}(2013){Jee}, {Tyson}, {Schneider}, {Wittman},
  {Schmidt}, \& {Hilbert}}]{Jee2013}
{Jee} M.~J., {Tyson} J.~A., {Schneider} M.~D., {Wittman} D., {Schmidt} S.,
  {Hilbert} S., 2013, \apj, 765, 74

\bibitem[{{Le F{\`e}vre} {et~al}\mbox{.}(2005){Le F{\`e}vre}, {Guzzo},
  {Meneux}, {Pollo}, {Cappi}, {Colombi}, {Iovino}, {Marinoni}, {McCracken},
  {Scaramella}, {Bottini}, {Garilli}, {Le Brun}, {Maccagni}, {Picat},
  {Scodeggio}, {Tresse}, {Vettolani}, {Zanichelli}, {Adami}, {Arnaboldi},
  {Arnouts}, {Bardelli}, {Blaizot}, {Bolzonella}, {Charlot}, {Ciliegi},
  {Contini}, {Foucaud}, {Franzetti}, {Gavignaud}, {Ilbert}, {Marano}, {Mathez},
  {Mazure}, {Merighi}, {Paltani}, {Pell{\`o}}, {Pozzetti}, {Radovich},
  {Zamorani}, {Zucca}, {Bondi}, {Bongiorno}, {Busarello}, {Lamareille},
  {Mellier}, {Merluzzi}, {Ripepi}, \& {Rizzo}}]{LeFevre2005}
{Le F{\`e}vre} O. {et~al.}, 2005, \aap, 439, 877

\bibitem[{Lemire \& Boytsov(2012)}]{Lemire2012}
Lemire D., Boytsov L., 2012, CoRR, abs/1209.2137

\bibitem[{{Mallat} \& {Zhang}(1993)}]{Mallat1993}
{Mallat} S.~G., {Zhang} Z., 1993, IEEE Transactions on Signal Processing, 41,
  3397

\bibitem[{{Mandelbaum} {et~al}\mbox{.}(2008){Mandelbaum}, {Seljak}, {Hirata},
  {Bardelli}, {Bolzonella}, {Bongiorno}, {Carollo}, {Contini}, {Cunha},
  {Garilli}, {Iovino}, {Kampczyk}, {Kneib}, {Knobel}, {Koo}, {Lamareille}, {Le
  F{\`e}vre}, {Le Borgne}, {Lilly}, {Maier}, {Mainieri}, {Mignoli}, {Newman},
  {Oesch}, {Perez-Montero}, {Ricciardelli}, {Scodeggio}, {Silverman}, \&
  {Tasca}}]{Mandelbaum2008}
{Mandelbaum} R. {et~al.}, 2008, \mnras, 386, 781

\bibitem[{{Myers} {et~al}\mbox{.}(2009){Myers}, {White}, \& {Ball}}]{Myers2009}
{Myers} A.~D., {White} M., {Ball} N.~M., 2009, \mnras, 399, 2279

\bibitem[{{Newman} {et~al}\mbox{.}(2013a){Newman}, {Cooper}, {Davis}, {Faber},
  {Coil}, {Guhathakurta}, {Koo}, {Phillips}, {Conroy}, {Dutton}, {Finkbeiner},
  {Gerke}, {Rosario}, {Weiner}, {Willmer}, {Yan}, {Harker}, {Kassin},
  {Konidaris}, {Lai}, {Madgwick}, {Noeske}, {Wirth}, {Connolly}, {Kaiser},
  {Kirby}, {Lemaux}, {Lin}, {Lotz}, {Luppino}, {Marinoni}, {Matthews},
  {Metevier}, \& {Schiavon}}]{Newman2013a}
{Newman} J.~A. {et~al.}, 2013a, \apjs, 208, 5

\bibitem[{{S{\'a}nchez} {et~al}\mbox{.}(2014){S{\'a}nchez}, {Carrasco Kind},
  {Lin}, {Abdalla}, {Amara}, {Banerji}, {Bonnett}, {Brunner}, {Carnero},
  {Castander}, {da Costa}, {Cunha}, {Fausti}, {Gerdes}, {Greisel}, {Gschwend},
  {Hartley}, {Jouvel}, {Lima}, {Maia}, {Marti}, \& {Miquel}}]{Sanchez2014}
{S{\'a}nchez} C. {et~al.}, 2014, in preparation

\bibitem[{{Sheldon} {et~al}\mbox{.}(2012){Sheldon}, {Cunha}, {Mandelbaum},
  {Brinkmann}, \& {Weaver}}]{Sheldon2012}
{Sheldon} E.~S., {Cunha} C.~E., {Mandelbaum} R., {Brinkmann} J., {Weaver}
  B.~A., 2012, \apjs, 201, 32

\bibitem[{{Wittman}(2009)}]{Wittman2009}
{Wittman} D., 2009, \apjl, 700, L174

\bibitem[{{York} {et~al}\mbox{.}(2000){York}, {Adelman}, {Anderson},
  {Anderson}, {Annis}, {Bahcall}, {Bakken}, {Barkhouser}, {Bastian}, {Berman},
  {Boroski}, {Bracker}, {Briegel}, {Briggs}, {Brinkmann}, {Brunner}, {Burles},
  {Carey}, {Carr}, {Castander}, {Chen}, {Colestock}, {Connolly}, {Crocker},
  {Csabai}, {Czarapata}, {Davis}, {Doi}, {Dombeck}, {Eisenstein}, {Ellman},
  {Elms}, {Evans}, {Fan}, {Federwitz}, {Fiscelli}, {Friedman}, {Frieman},
  {Fukugita}, {Gillespie}, {Gunn}, {Gurbani}, {de Haas}, {Haldeman}, {Harris},
  {Hayes}, {Heckman}, {Hennessy}, {Hindsley}, {Holm}, {Holmgren}, {Huang},
  {Hull}, {Husby}, {Ichikawa}, {Ichikawa}, {Ivezi{\'c}}, {Kent}, {Kim},
  {Kinney}, {Klaene}, {Kleinman}, {Kleinman}, {Knapp}, {Korienek}, {Kron},
  {Kunszt}, {Lamb}, {Lee}, {Leger}, {Limmongkol}, {Lindenmeyer}, {Long},
  {Loomis}, {Loveday}, {Lucinio}, {Lupton}, {MacKinnon}, {Mannery}, {Mantsch},
  {Margon}, {McGehee}, {McKay}, {Meiksin}, {Merelli}, {Monet}, {Munn},
  {Narayanan}, {Nash}, {Neilsen}, {Neswold}, {Newberg}, {Nichol}, {Nicinski},
  {Nonino}, {Okada}, {Okamura}, {Ostriker}, {Owen}, {Pauls}, {Peoples},
  {Peterson}, {Petravick}, {Pier}, {Pope}, {Pordes}, {Prosapio},
  {Rechenmacher}, {Quinn}, {Richards}, {Richmond}, {Rivetta}, {Rockosi},
  {Ruthmansdorfer}, {Sandford}, {Schlegel}, {Schneider}, {Sekiguchi}, {Sergey},
  {Shimasaku}, {Siegmund}, {Smee}, {Smith}, {Snedden}, {Stone}, {Stoughton},
  {Strauss}, {Stubbs}, {SubbaRao}, {Szalay}, {Szapudi}, {Szokoly}, {Thakar},
  {Tremonti}, {Tucker}, {Uomoto}, {Vanden Berk}, {Vogeley}, {Waddell}, {Wang},
  {Watanabe}, {Weinberg}, {Yanny}, {Yasuda}, \& {SDSS
  Collaboration}}]{York2000}
{York} D.~G. {et~al.}, 2000, \aj, 120, 1579

\end{thebibliography}

\bsp
\label{lastpage}
\end{document}